\documentclass[conference]{IEEEtran}
\IEEEoverridecommandlockouts
\usepackage{cite}
\usepackage{amsmath,amssymb,amsfonts}
\usepackage{algorithmic}
\usepackage{graphicx}
\usepackage{textcomp}
\usepackage{xcolor}
\usepackage{balance}
\usepackage{url}
\def\BibTeX{{\rm B\kern-.05em{\sc i\kern-.025em b}\kern-.08em
    T\kern-.1667em\lower.7ex\hbox{E}\kern-.125emX}}
\begin{document}

\title{Automating Static Code Analysis Through CI/CD Pipeline Integration}

\makeatletter
\newcommand{\linebreakand}{%
  \end{@IEEEauthorhalign}
  \hfill\mbox{}\par
  \mbox{}\hfill\begin{@IEEEauthorhalign}
}
\makeatother

\author{\IEEEauthorblockN{Zachary Wadhams}
\IEEEauthorblockA{\textit{Gianforte School of Computing} \\
\textit{Montana State University}\\
Bozeman, Montana, USA \\
zacharywadhams@montana.edu}
\and
\IEEEauthorblockN{Ann Marie Reinhold}
\IEEEauthorblockA{\textit{Gianforte School of Computing} \\
\textit{Montana State University}\\
Bozeman, Montana, USA \\
reinhold@montana.edu}
\and
\IEEEauthorblockN{Clemente Izurieta}
\IEEEauthorblockA{\textit{Gianforte School of Computing} \\
\textit{Montana State University}\\
\textit{Pacific Northwest National Laboratory}\\
\textit{Idaho National Laboratory}\\
Bozeman, Montana, USA \\
clemente.izurieta@montana.edu}
}

\maketitle

\begingroup
\renewcommand\thefootnote{}
\footnotetext{%
\footnotesize
© 2024 IEEE. Personal use of this material is permitted.
Permission from IEEE must be obtained for all other uses, in any current or future media,
including reprinting/republishing this material for advertising or promotional purposes,
creating new collective works, for resale or redistribution to servers or lists,
or reuse of any copyrighted component of this work in other works.

\medskip
This is the author's accepted version of:
Z. Wadhams, A. M. Reinhold, and C. Izurieta,
``Automating Static Code Analysis Through CI/CD Pipeline Integration,''
2024 IEEE International Conference on Software Analysis, Evolution and Reengineering
(SANER), Rovaniemi, Finland, 2024, pp. 119--125.
doi: \url{https://doi.org/10.1145/3691621.3694947}.
}
\addtocounter{footnote}{-1}
\endgroup

\begin{abstract}
In the contemporary landscape of software development, securing sensitive data is paramount to safeguarding organizational reputation, preventing financial losses, and protecting individuals from identity theft. This paper addresses the pervasive challenge of identifying and rectifying security vulnerabilities early in the development process, emphasizing the role of Static Application Security Testing (SAST) tools. While SAST tools play a crucial role in detecting vulnerabilities, widespread adoption has been hindered by usability issues, including high false positive rates and a lack of native pipeline support. This paper proposes a novel, generalized, and automated process for aggregating SAST tool outputs and integrating them into developers' familiar issue-tracking software. The process streamlines the identification and communication of security vulnerabilities during the development lifecycle, facilitating more efficient remediation efforts. We demonstrate the successful implementation of the proposed process with the SonarQube SAST tool in a GitLab-based development environment. Developers were positive about the structured implementation, real-time feedback, and proactive vulnerability management. However, despite some challenges such as a potential learning curve and trade-offs between secure coding and workflow disruption, the overall positive impact on security awareness and responsiveness suggests that the proposed process holds promise in enhancing the security posture of software development practices
\end{abstract}


\section{Introduction}
Now, more than ever, software applications must make a concerted effort to effectively secure the data they store. A single breach of security can wreak havoc on the reputation of the organization, trigger massive financial loss, and even disrupt individuals' lives through identity theft or other means. Many instances of security breaches can be traced back to security vulnerabilities \cite{b7}. A security vulnerability is a weakness or flaw in a software application's source code or design that can be exploited by malicious actors to compromise the security of the system or the data it processes. It is vital that vulnerabilities are identified and corrected as early as possible in the development process. A key tool that development teams can use to identify vulnerabilities is Static Application Security Testing (SAST), or Static Code Analysis. 

These tools aim to help ensure the security, reliability, and compliance of software applications. SAST tools work by analyzing the source code during the development phase, enabling early detection of security vulnerabilities. By identifying security flaws at an early stage, developers can address them promptly and minimize the risk of such vulnerabilities making their way into the final product. 

SAST tools not only focus on security vulnerabilities but also help improve code quality by identifying design flaws \cite{b1} \cite{b2}. By detecting and addressing these issues early in the development cycle, teams can enhance the overall quality and maintainability of the codebase. Using these tools, development teams can gain insights into common security pitfalls and improve their understanding of secure coding principles. SAST tools simultaneously serve as educational resources, helping to foster a security-conscious culture among developers who utilize these tools and empowering them to produce code that is more secure and reliable \cite{b5}. 

A small number of such tools are specifically designed to be implemented within Continuous Integration/Continuous Delivery(CI/CD) pipelines on major Git-based repository hosting services. CI/CD pipeline integration allows these tools to run on each new code addition, creating a record (typically stored within the tool GUI or as a .json file) of which changes introduced vulnerabilities or design flaws. The near-immediate feedback given allows for faster remediation of critical vulnerabilities while the historical record can be instrumental in ensuring compliance with a variety of security standards and best practices \cite{b11} \cite{b12}. However, the vast majority of tools have little to no support for automation or pipeline integration.

While manual code reviews and standalone static analyses are essential for identifying some design flaws and guiding a project's direction, the complexity, and size of modern software applications make it unfeasible to manually review an entire codebase \cite{b9}. This necessity calls for automation, which is provided by pipeline integration.

Despite these apparent benefits, many development teams have not deployed SAST tools. Reasons for this vary but many agree that there are still widespread usability issues that hinder their adoption and consistent use. Most of these reasons are commonly identified as high rates of false positives, unhelpful warning messages, lack of fix suggestions, and insufficient native pipeline support \cite{b3}. Developers often delegate security-related concerns to colleagues within their organization or, in some cases, they may lack access to these reports due to security configurations \cite{b8}\cite{b4}. As those possessing the most profound understanding of the code, developers often overlook security issues, which can lead to the ``out of sight, out of mind" problem \cite{b8} \cite{b5}.

Instead of merely investigating this issue, we have chosen to design a solution that involves presenting SAST reports to developers in a familiar and consistent manner using automated methods. While many have touched upon this issue, we have identified no papers that discuss a generalized process applicable to all static analysis tools. By offering a standardized process that is easy to integrate and demonstrating the tangible benefits of doing so, we aspire to promote the widespread adoption and integration of static analysis tools within the software development community. This, in turn, enhances software privacy and security across a broad spectrum of domains and applications. 

Our first key contribution is a \textbf{generalized and automated} process for aggregating SAST tool outputs and integrating them into developers' familiar issue-tracking software. This process streamlines the identification and communication of security vulnerabilities during the development lifecycle, facilitating more efficient remediation efforts.

In addition, we provide a detailed examination of a practical implementation of the proposed process. By illustrating its application in a real-world scenario, we offer insights into the feasibility and effectiveness of the process. The practical implementation serves as a valuable case study, shedding light on the challenges encountered and the practical considerations that arise when implementing such a process within software development teams.

Our work contributes to the security vulnerability management field by thoroughly exploring the potential benefits and drawbacks of the explored process. By addressing advantages such as improved collaboration between security and development teams, which in many cases are separated by technology and/or organizational constraints, our aids in understanding the implications and trade-offs associated with adopting our process.

Importantly, these contributions hold direct relevance to the Mining Software Repositories (MSR) community. The proposed process aligns with the objectives of MSR by leveraging data from software repositories, issue-tracking systems, and security testing tools to better inform software developers of potential security vulnerabilities and foster proactive measures for secure software development practices, ergo the ability to mine software repositories to make information available to potentially separate engineering groups (i.e., security and development) in a CI/CD environment is critical. The insights gained from this research can inform and enhance the broader understanding of how security practices intersect with the larger software development ecosystem, thereby contributing to the ongoing advancements in the MSR field.

This paper is structured as follows: Section \ref{sec:Motivation} explains the motivation for our work in the field of static analysis. Section \ref{sec:Related Work} discusses related work and its impact on our research. Section \ref{sec:Process} describes our process and its core components in detail. Section \ref{sec:Use Case} presents a practical example of an implementation of our process. Section 
\ref{sec:Discussion} provides a discussion on the greater consequences of our work. Section \ref{sec:Threats to Validity} addresses potential threats to the validity of our work and our attempts to mitigate them. Section \ref{sec:Conclusion and Future Work} concludes the paper and outlines our plans for future work on refining and extending the proposed SAST tool integration process.

\section{Motivation}
\label{sec:Motivation}
It is the moral and now legal obligation of organizations to ensure that any and all software they release is as safe and secure as possible and respects the privacy of its users. The EU's General Data Protection Regulation (GDPR) is just one example of legislation which requires that organizations ensure privacy and security are built into their applications by design \cite{b10}.

SAST tools excel in this area where other methods, such as dynamic analysis, fall behind. Dynamic analysis is performed later in the development process, typically when an application is executed. Static analysis strictly looks at source code and can be implemented as soon as the first line of code is written. This reinforces SAST's use as a proactive step that can be used early in the development process when an application's design is still flexible, bringing it in line with the core tenet of privacy and security by design. Although static analysis procures many benefits, it is often utilized inconsistently.

Issues with organizational security configurations can make the consistent use of SAST difficult for some. Occasionally, organizations restrict the execution of tools, requiring them to be behind a firewall and rendering them inaccessible to certain developers. This becomes a problem when those with access are senior developers who do not have time to sift through the reports generated by tools, leading to slow turnaround times on fixes or even vulnerabilities being missed completely. Thus, a disconnect exists between the development teams and security experts or senior developers who are directly responsible for ensuring code security.

This disconnect demands a better way to bring more average developers into the sphere of security and subsequently motivated our team to design a process that does just that. Leveraging the familiarity of bug and issue-tracking features within popular repository hosting services, the goal is to increase the visibility of SAST tool outputs, ultimately providing a consistent and accessible space for identifying, managing, and resolving design flaws and vulnerabilities within software repositories. 

\section{Related Work}
\label{sec:Related Work}
Interest in SAST has been steadily growing over the past two decades, while interest in security and privacy by design has burgeoned since the publication of the GDPR in 2016 \cite{b10}. The combination of SAST tools and the concept of privacy and security by design, and how one can drive the usage of the other, inspired our work on the topic.

Johnson et al. \cite{b13} carried out interviews to investigate why static analysis tools were not being used by developers. During their research, they uncovered many underlying issues with these tools, such as poor output presentation and slow feedback.  In other work performed by Izurieta et al. \cite{b16}, the uncertainty of scoring and error propagation from SAST tools is also addressed.

Haug et al. \cite{b11} showed that when developers are provided with immediate or near-immediate feedback on code, they are more likely to consider it.

Xie et al. \cite{b4} found that most software security vulnerabilities are caused by errors introduced by developers. Their results led them to discover a striking divide between developers' security knowledge and their practices. They concluded that static analysis tools do play an important role in assisting developers in producing secure software and overcoming their apprehension towards security.

Ayewah et al. \cite{b3} observed that when developers are present with security-related messages, they generally make the correct decisions to address them even with little to no formal security experience.

Nachtigall et al. \cite{b5} conducted a comprehensive study focused on specific criteria for static analysis tools. They analyzed 36 criteria across 46 different tools from a user's perspective. The study revealed significant shortcomings in many SAST tools, particularly concerning their integration into developers' workflows. It was identified that the locations where tool outputs are stored is unfamiliar to many developers and accessing them often requires them to change their habits. In some cases, these tools have their data stored behind firewalls that only senior developers or cybersecurity officers can access. Whether intentional or not, this leaves other developers in the dark about the potential vulnerabilities they may introduce, thereby compromising the security culture of the organization. Many tools also provide these outputs in unformatted text that doesn't grab a developer's attention. As a result, they found that if a tool disrupts a developer's workflow or lacks sufficient visual guidance, many will quickly abandon it. This underscores a critical and recurring usability challenge. 

Our process not only addresses these previously identified issues but also establishes a way for delivering static analysis tool outputs to developers in a non-disruptive and visually appealing format, ultimately ensuring that a broader range of developers can easily access these reports, thereby enhancing, rather than impeding, their productivity and workflow.

\section{Process}
\label{sec:Process}
We implement our process in three steps First, we identify the SAST tools preferred by the organization and assess the types of data they provide. Two, we explore the development environment of the target organization, taking into account aspects such as issue-tracking software, repositories, pipeline configuration, and network security configurations. Three, we implement a controller script that ties the SAST tools and issue tracking together. The sub-sections below provide a detailed explanation of each of these steps.

\begin{figure}[t]
\centerline{\includegraphics[scale=0.40]{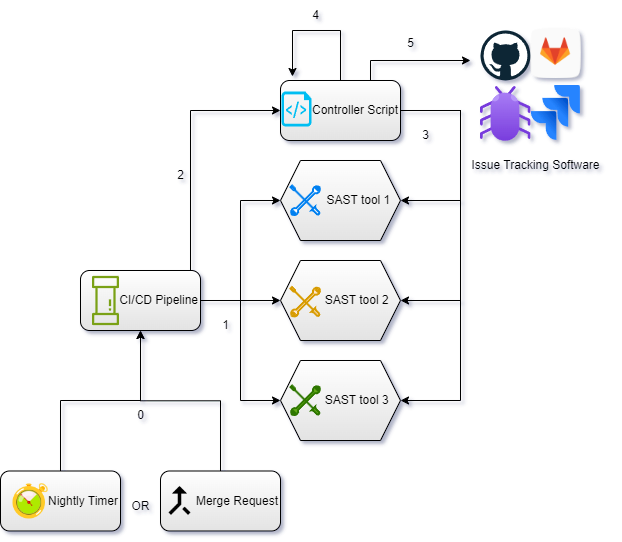}}
\caption{ Process flow diagram. \\ \\ The boxes represent important concepts or technologies while the arrows depict the flow of the process. In step \textbf{0}, some external factor, such as a nightly timer or a developer-initiated merge request, triggers the build pipeline. The pipeline then initiates the analysis of each SAST tool in step \textbf{1}. Once all static analyses are completed, the build pipeline starts the controller script in step \textbf{2}. The controller script reaches out to each SAST tool and gathers the relevant issue data in step \textbf{3}. The issue data is then formatted by the controller script and assembled into payloads in step \textbf{4}. In step \textbf{5}, each payload is sent to the issue tracking software, and individual issues are created.}
\label{figure1}
\end{figure}

\subsection{Tool Identification and Data Assessment}
Diverse organizations have unique analytical requirements, prompting the need for discussions on preferred static analysis tools. These tools can be categorized into two groups: those with a Graphical User Interface (GUI) and those with a Command Line Interface (CLI). Our process focuses on GUI tools. They typically have the most customizability and often offer an Application Programming Interface (API). The API simplifies data retrieval by returning it seamlessly in commonly formatted structures such as XML, JSON, or HTML. Our API-driven process enhances the efficiency of data extraction, making it easier to integrate the data into our controller script. 

In our context, we treat GUI tools without APIs the equivalently to CLI tools. CLI tools are executed exclusively from the command line and do not provide the additional functionality and ease of use offered by an API. After completion, these tools generate a file formatted as either XML or JSON containing the report data. While CLI tools can provide valuable information, our process does not focus on them.

After identifying which tools the organization desires, we conducted an investigation of the organization's familiar development environment.

\subsection{Development Environment Exploration}
The development environment of the target organization is taken into account when custom tailoring the implementation of our process. We considered what software most organizations use to manage their codebase and found that the vast majority use one of three Git-based systems; GitHub, GitLab, and BitBucket. Many development teams either utilize the built-in repository issue tracking or rely on connected software adjacent to the repository for issue management \cite{b14}. Each of these services provides a robust API to assist with automating aspects such as bug and issue tracking. These APIs allow for calls to be made that create, edit, or resolve issues. An ``issue" or sometimes ``ticket" is a digital record used to track tasks, bugs, and feature requests related to a software project. They help teams collaborate by providing a place to discuss, assign, and monitor the progress of these work items. APIs are ubiquitous in issue tracking, enabling this process to be applied to any service through the controller script.

The build pipeline of the repository is a predefined and automated sequence of tasks and actions that transform source code into a deployable application or software artifact. This pipeline is where the code is compiled and also the stage at which SAST tools are executed. The CI/CD pipelines of the aforementioned repository hosting services are standardized through the use of a common file type used to choreograph the execution steps of the pipeline. The controller script is placed after the SAST tool execution in the pipeline order to ensure that each analysis is completed before any data is retrieved. 

As previously mentioned in section II (i.e., Motivation), some organizations may have security configurations that restrict the execution of tools or the CI/CD pipeline behind a firewall, rendering them inaccessible to certain developers. Our process accommodates such security controls while allowing for developer access, as long as the machines running the tools and the pipeline can be configured to communicate with each other and bypass firewall rules. Some organizations may cite vulnerability data as a potential security concern that should not be shared widely. However, the developers who would be fixing security vulnerabilities must already have access to the project's source code. This source code access, combined with the fact that many SAST tools are open source and free to use, means that anyone with source code access could run an analysis of their own and obtain these reports if they wished to, making the reasoning behind this security concern unsound.

After collecting the necessary tools and information about the development environment, the controller script can take shape.

\subsection{Controller Script}
\label{sec:Controller Script}
The controller script serves as the keystone that seamlessly connects the previously disjoint processes of SAST report generation and issue management. Any scripting language is acceptable as long as it is supported by the hosting pipeline. Its primary role is to orchestrate the flow of data between the SAST tools and the issue tracking system, ensuring a smooth and automated transition. This bridge is established through a series of well-defined steps. 

The controller script begins by collecting the SAST report data generated by whatever tools the target organization chooses to use (Figure \ref{figure1}, step 3).  Through the tool's API, raw SAST data is obtained, and the controller script transforms it into a standardized format that aligns with the requirements of the issue tracking system (Figure \ref{figure1}, step 4). At a minimum, the format includes a title that identifies the issue type and severity, a one-sentence description of the problem, an identifier indicating the specific line of code and file where the issue resides, and either a problem description if provided by the tool or a unique reference number such as Common Vulnerabilities and Exposures (CVE) or Common Weakness Enumeration (CWE). This transformation ensures that data from diverse tools is uniform and can be consistently integrated into the tracking system. Before sending the data as an issue to the tracking software, it must be converted to markdown, ensuring compatibility and consistency with the tracking system's formatting and requirements.

Leveraging the issue tracking system's API  the controller script automates the creation (Figure \ref{figure1}, step 5) of issues or tickets to report vulnerabilities, weaknesses, or other code-related concerns identified by the SAST tools. It establishes near real-time synchronization through the pipeline, updating the issue tracking system with the latest analysis results whenever the pipeline runs due to developers committing code. (Figure \ref{figure1}, step 0) 

To prevent the creation of duplicate issues within the issue tracking software, issues within the SAST tool must be marked. One straightforward process is to update the status of all issues that the script has identified and 'moved' within the tool. This ensures that upon each new analysis, issues that have already been seen by the script are ignored and only new issues are considered (Figure \ref{figure1}, step 3). 

In many cases, it may be valuable for the script to include a quality gate. Quality gates play a crucial role in maintaining the integrity of the software development process. They function by continuously monitoring the code changes as they pass through the CI/CD pipeline. If the gate detects the introduction of major vulnerabilities or other high-risk issues, it will immediately halt the pipeline's progress after generating the issues, alerting developers of a potential problem. This preventive measure ensures that no changes with severe flaws are allowed to proceed further into the development or deployment stages.

In practice, a quality gate serves as an additional layer of defense, reinforcing the security and quality of the software. It ensures that any code changes are thoroughly examined for major vulnerabilities before they can proceed, thereby contributing to a more robust and reliable software development process.

\begin{figure}[t]
\centerline{\includegraphics[scale=0.35]{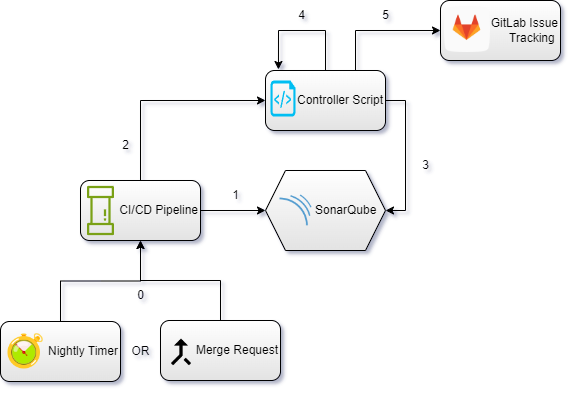}}
\caption{Implementation of Approach to an Example Organization (Organization X). \\ \\ The boxes represent important concepts or technologies while the arrows depict the flow of the process. The implementation begins with either a nightly timer or merge request in step \textbf{0}. The pipeline then initiates the analysis SonarQube in step \textbf{1}. Once SonarQube's analysis is complete, the build pipeline starts the controller script in step \textbf{2}. The controller script reaches out to SonarQube and gathers the relevant issue data in step \textbf{3}. The issue data is then formatted by the controller script and assembled into payloads in step \textbf{4}. In step \textbf{5}, each payload is sent to GitLab's issue tracking software, and individual issues are created therein.}
\label{figure2}
\end{figure}

The versatility of the controller script is a fundamental strength of our process, as it can be tailored to collect diverse types of issues such as vulnerabilities, bugs, or design issues while filtering them by potential severity or impact. The script can be configured to recognize and process different issue categories by adjusting the data collection and transformation steps, thereby accommodating the specific needs and priorities of the development team. By integrating these functions, the controller script effectively bridges the gap between the security-focused static analysis and the broader software development process, fostering an efficient and proactive process to addressing code vulnerabilities and design concerns.

\section{Use Case}
\label{sec:Use Case}
To test the practical application of our process we worked with an organization whose goal was to implement a static analysis tool to enhance the security of their in-development application. To protect this organization's identity, we refer to them as Organization X. 

Figure \ref{figure2} depicts Organization X's implementation of our process, following the same steps outlined in Figure \ref{figure1}.

Being the first static analysis to be used on their repository, a tool that is simple to configure but still offers customizability and scales well with a rapidly growing codebase was deemed necessary. Following our outlined process, Organization X selected SonarQube as their preferred tool after evaluating its features and usability. 

The development environment of Organization X consisted of a GitLab repository where code is hosted, the pipeline is managed, and issues are tracked (Figure \ref{figure2}, steps 0, 5). A key facet to note is that the repository build pipeline as well as all execution of external tools were required to be placed behind a firewall to comply with the organization's security requirements. This was a hard requirement that, as a consequence, restricted who could work with SonarQube to two senior developers out of more than twenty total developers. As SonarQube stores outputs within its GUI, this issue exemplifies the problem discussed in section II (i.e., Motivation) where some developers are segregated due to accessibility of sensitive data. In the absence of a structured process, only two developers would be able to configure, maintain, and then be responsible for the potentially very large number of vulnerabilities and design flaws that could be uncovered. 

\begin{figure}[t]
\centerline{\includegraphics[scale=0.35]{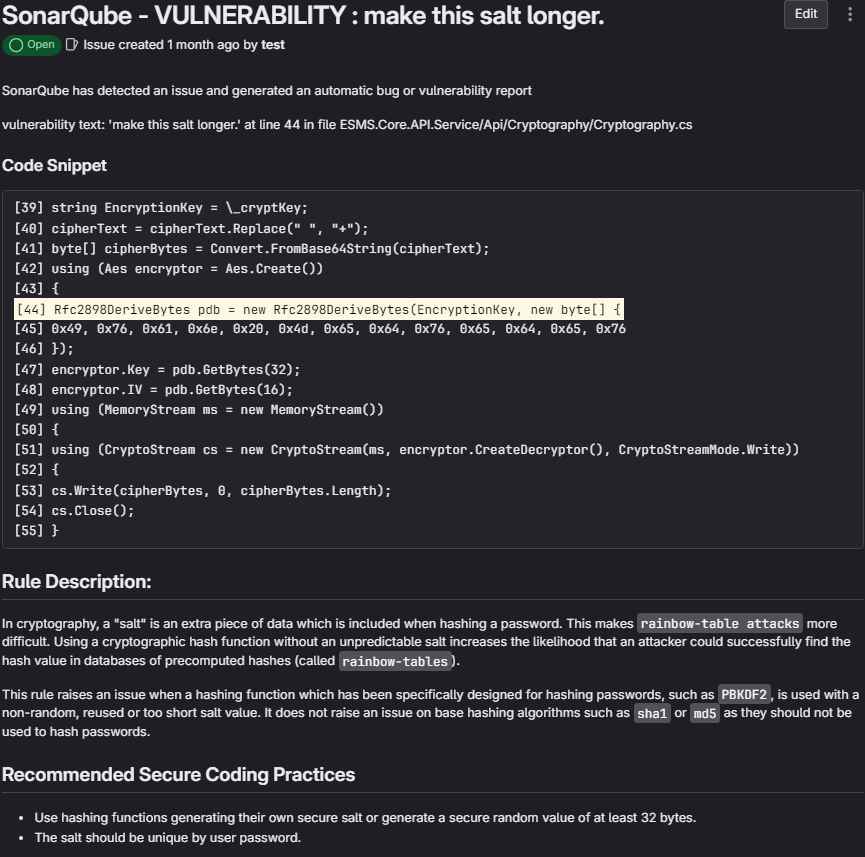}}
\caption{Example of a Generated Issue}
\label{figure3}
\end{figure}

Organization X used our process, implementing the controller as a Python script (Figure \ref{figure2}, step 2). This script utilized the Requests package to make HTTPS requests to the SonarQube and GitLab instances. SonarQube has the capability to uncover three different types of issues in projects: bugs, vulnerabilities, and code smells. Organization X was interested only in vulnerabilities and bugs, so these requests were targeted to endpoints within SonarQube's Web API to retrieve data related to those types of issues (Figure \ref{figure2}, step 3). SonarQube also attaches a severity to each issue it finds. Organization X decided to consider only bugs and vulnerabilities with a severity of critical, major, or high in an effort to triage issues that have a larger potential to cause problems. 

After retrieving the data, the script extracted the essential information from the JSON objects. Subsequently, a payload was assembled and sent to GitLab's issue-tracking software (Figure \ref{figure2}, step 4). Organization X decided on the data that would be most relevant and useful to their developers, choosing to include selected information in generated issues. The one-sentence description of the created GitLab issue comprised the SonarQube issue type, title, and severity. The body of the issue is generated with a disclaimer, stating that this issue was automatically created using data from SonarQube. Organization X considered this necessary to help their developers differentiate between automatically generated and manually created issues. The rest of the issue body contained a code snippet that showed the problem line of code along with the surrounding 10 lines for context, an explanation of why the issue should be addressed, and a link to the corresponding CWE or CVE for further reading. Each payload was then sent to GitLab's API as a create issue request and would appear alongside developer-created issues within GitLab's issue tracking software for developers to pick up and address (Figure \ref{figure2}, step 5).

As this process was tested, starting with Organization X's nightly pipeline runs, they found it to be helpful as it didn't disrupt their established pipeline and did not add a significant amount of runtime. Developers also reacted positively to the formatting and location of the issues, commenting that the problems in the code were easily identifiable, and the additional information provided aided them in engineering fixes. Figure \ref{figure3} depicts an example issue that Organization X generated in GitLab using data from SonarQube.

After testing for a few months and being satisfied with the results, Organization X decided to develop a secondary controller script designed to run on their merge request pipeline with every developer code change. The aim of this was to prevent developers from introducing vulnerabilities and bugs through the addition of a quality gate. When a developer attempts to merge their code with the main code branch, the quality gate checks whether the code changes would introduce bugs or vulnerabilities. If they do, the changes are rejected, and the developer is notified to fix the issues before merging their code. If the changes do not introduce bugs, the merge proceeds as normal. Organization X chose to reject changes if they introduce more than one blocker vulnerability or bug or more than two critical vulnerabilities or bugs.

\section{Discussion}
\label{sec:Discussion}
The developer feedback provided us with insights into how they viewed different aspects of the implementation of SAST tools into their workflow using our process. For example, when the quality gate was implemented, a few violations were noted on the first day of use. The gate functioned as expected, rejecting the changes. Feedback on the gate from developers was conflicting; while all understood its necessity, opinions varied on its rigidity. Some opposed the aggressive process, suggesting issues be noted for later resolution to allow focused coding. Others appreciated it, foreseeing time saved in the long run. This pinpointed the quality gate as a potential challenge in terms of the developer's workflow and productivity.

Developers appreciated the structured implementation of the process, expressing satisfaction with the automation of vulnerability and bug identification while positive sentiments were shared about the seamless integration with GitLab's API and the existing pipelines. Perhaps the most commonly identified upside to the automated generation of issues was the real-time feedback, which either helped developers immediately fix the issues or begin planning to address them at a future date. Developers acknowledged the significant enhancement in code security through systematic vulnerability identification and positively recognized the proactive process of the quality gate in effectively managing vulnerabilities. Over a short period, numerous previously unknown issues were uncovered. While some of these turned out to be false positives, others were genuine and had been lurking within the codebase for months. Organization X's developers stated that without the implementation of static analysis, these issues might have gone undetected. As developers are exposed to more vulnerabilities, there is potential for a relief of tension between development and security teams through increased collaboration. 

However, our process was not without complications. A potential problem developers identified was that due to the formatting, identifying false positives required more effort. Developers had to delve deeper into an issue before realizing that it was a non-issue. This could be argued as a trade off for having issues formatted in a detailed way. The introduction of the new process faced initial resistance as it presented a learning curve to developers who were already comfortable with existing practices.

The developer feedback we received illuminates important questions to the MSR community. What is the value of the trade-off between having secure code and the potential disruption to a developer's workflow? How can a balance between those two important aspects be achieved? It is evident that while the benefits of the automated generation of issues are substantial, addressing the initial resistance and facilitating a seamless adaptation process are key aspects to consider in the ongoing refinement of the process. 

\section{Threats to Validity}
\label{sec:Threats to Validity}
Our study contributes valuable insights to the fields of SAST, vulnerability management, and MSR. Here we address potential threats to the validity in our study. By identifying and acknowledging these limitations, we aim to provide a transparent assessment of the scope and generalizability of our work. This section outlines key threats to the validity of our study, offering a comprehensive view of the factors that may impact the reliability and applicability of our results.

First, while we attempted to make our process as generalized as possible, Organization X may have unique characteristics that limit the generalizability of findings to other organizations. To address this, we thoroughly explored their development environment and compared it to other known development environments. While hardly exhaustive, we identified no known out-of-the-ordinary development practices. Second, the definition and identification of vulnerabilities and bugs may vary among different SAST and their versions \cite{b15}. Since SonarQube was the sole tool utilized by Organization X, we lack a practical implementation of other tools to compare our results. Addressing this limitation is part of our future work, as detailed in the following section, where we plan to test with additional tools to confirm the generalizability of our process.

\section{Conclusion and Future Work}
\label{sec:Conclusion and Future Work}
In this paper, we have addressed the critical issue of effectively integrating SAST tools into the software development lifecycle. This work's noteworthiness to the MSR community cannot be understated due to its direct alignment with the overarching goals of deriving actionable information relating to privacy and security. The increasing importance of securing data and the potential consequences of security breaches highlight the necessity for proactive measures in identifying and addressing security vulnerabilities. Our process focuses on automating the aggregation and integration of SAST tool outputs into developers' familiar issue-tracking software, thereby enhancing the visibility and accessibility of security-related issues. Our results from the use case suggest that developers are mostly satisfied with the way these issues are presented and that the addition of them to their workflow has not been overly intrusive.

While our results are encouraging, there is still much work to be done on the our procedure, such as adding additional functionality and conducting use cases with other tools. For instance, many SAST tools are CLI-only and lack an API for data retrieval. In the future, we will address these tools as well by incorporating them into our process, rather than solely focusing on large web-based tools. Some of these CLI tools provide little to no detail on the issues they uncover, except for a reference to their corresponding CWE or CVE. We will investigate whether retrieving data from the CWE/CVE databases to fill out more information in these issues would be valuable. Additionally, we plan to conduct other use cases that follow our process with different SAST tools, potentially incorporating more than one tool at once. 

\section{Acknowledgements}
This research is supported by TechLink (TechLink PIA FA8650-23-3-9553). The ChatGPT Large Language Model was used in this paper for spell-checking and grammatical enhancements.

\balance

\vspace{12pt}

\end{document}